\documentclass[a4paper]{jpconf}
\usepackage{graphicx}
\usepackage{amsmath}
\usepackage{amssymb}
\usepackage{xcolor}
\usepackage{xcolor}
\usepackage{xcolor}
\usepackage{setspace}
\usepackage{cases}
\usepackage{bm}
\usepackage{enumitem}
\usepackage{booktabs}
\usepackage{textcomp}
\usepackage{makecell}
\usepackage{multirow}
\usepackage{wrapfig, blindtext}

\begin{document}
\title{Excitation of a homogeneous dielectric sphere \\by a point electric dipole}

\author{Roman Gaponenko,$^{1,*}$ Ilia L. Rasskazov,$^2$ Alexander Moroz,$^3$  \\Dmitry Pidgayko,$^1$ Konstantin Ladutenko,$^1$ Alexey Shcherbakov,$^1$ and Pavel Belov$^1$}

\ead{roman.gaponenko@metalab.ifmo.ru}

\address{
 $^1$ School of Physics and Engineering, ITMO University, 197101, Saint-Petersburg, Russia \\ $^2$ The Institute of Optics, University of Rochester, Rochester, New York 14627, United States \\
 $^3$ Wave-scattering.com
}

\begin{abstract}
Electrically small dielectric antennas are of great interest for modern technologies, since they can significantly reduce the physical size of electronic devices for processing and transmitting information.
We investigate the influence of the resonance conditions of an electrically small dielectric spherical antenna with a high refractive index on its directivity and analyze the dependence of these resonances on the effectively excited modes of the dielectric sphere. 
\end{abstract}

\section{Introduction}
\label{sec:intro}

The first antennas based on the use of a dielectric resonator appeared at the end of the last century. They have recently become in high demand during the rapid development of communication networks. 
This process involves a gradual increase in the quality and speed of information transfer, which inevitably forces telecom operators to move to higher frequencies, requires a reduction of the physical dimensions of devices, and also a minimization of interference effects. The development of super directional antennas can be essential for a broad variety of applications.

Dielectric resonator antenna (DRA) is currently one of the most attractive antenna types for use in the microwave and millimeter wave bands due to simplicity of design, small physical size, wide frequency range (0.7...40GHz), high radiation efficiency, varying polarizations, multiple feed systems, stable radiation patterns, ease of integration with other antennas, and high temperature
\begin{wrapfigure}{r}{5cm}
\centering
\includegraphics[width=3.5cm]{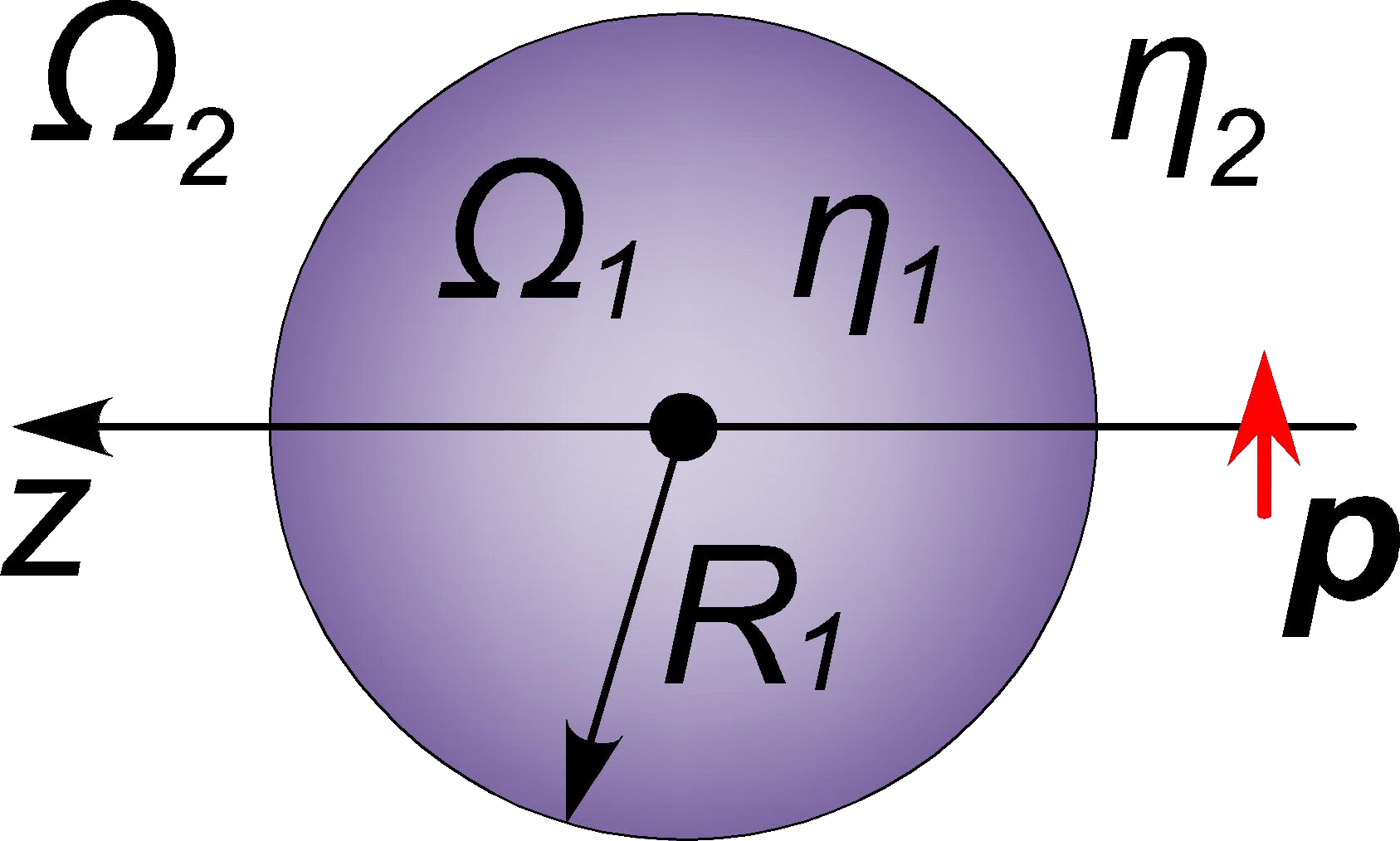}
\caption{\label{fig:sphere}
Sketch of the problem under study.}
\vspace{-10mm}
\end{wrapfigure} 
tolerance.
The absence of metal parts makes it possible to work more efficiently than metal antennas at high frequencies, since replacing conductive materials with dielectrics eliminates ohmic losses, as well as the lack of additional power loss into their dielectric substrates through surface waves. 
The absence of any metal components makes such DRA's attractive for space applications.
DRA's have also its disadvantages, such as low mechanical strength (for most materials used), sensitivity to temperature changes and mechanical vibrations.

\section{Formulation}
\label{sec:optim}

DRA's are known to have resonant behavior \mbox{\cite{Debye1908, Richtmyer1939,Stratton_1941, Gastine1967}}. Resonances in scattering correspond to the poles of the scattering coefficients \cite{Bohren1998}. They are often called ``morphology-dependent resonances'', which emphasizes their dependence on the parameters of the sphere, such as the refractive index and the ratio of the radius of the sphere to the wavelength. 
From a physical point of view, the resonances are directly related to an efficient interaction of an incident electromagnetic wave with the sphere through the excitation the sphere's eigenmodes \cite{Forestiere2019,Gaponenko2021}. In order to achieve a directional scattering, it is necessary that the emission frequency of an electric dipole is close to one of the resonant frequency. 
Usually, to obtain a directional radiation in one direction, it is considered sufficient to suppress radiation in the opposite direction using the interference of electric and magnetic modes of the scatterer \cite{Paniagua-Dominguez2019}. 
While this method allows to get directional designs, it does not allow to always achieve the highest possible directivity, especially for electrically small DRA's with a high refractive index \cite{Gaponenko2021}.

We consider an electrically small lossless antenna in the form of a homogeneous dielectric sphere located in air and excited by a {\em tangentially} oriented point electric dipole $\textbf{p}$ as shown in Figure \ref{fig:sphere}. 
The dipole can be located both inside and outside DRA. 
The center of coordinates is at the sphere origin. 
We use the following notation here: the dipole position, $r_d$; radius of the spherical dielectric antenna, $R_1$; a homogeneous isotropic concentric domain, $\Omega_1$; refractive index of the sphere, $\eta_1 \geq 1$; refractive index of the surrounding medium, $\eta_2=1$ (in the domain $\Omega_2$). 
We consider homogeneous, isotropic, non-magnetic materials with the permeability of a vacuum, $\mu_0$. 
It is worth noting that for a resonant condition, especially for a homogeneous sphere with a high refractive index, irregularities such as roughness and bulge can be important, but in this work the case of a perfectly smooth surface is assumed. 

In what follows, we make use of an exact analytical solution to the problem of scattering of an electric dipole radiation on a sphere~\cite{moroz_recursive_2005}. The resonance conditions can be calculated from the poles of the scattering coefficients \cite[eqs.(63),(66)]{moroz_recursive_2005}, which, under some circumstances, can be similar to the conditions for electromagnetic resonances of free dielectric sphere \cite{Gastine1967}:
\begin{align}
\begin{aligned}
    &\text{Mode $TE_{lms}$:\hspace{0.5cm}} {j_{l-1}(\eta_1 kR_1)}/{j_{l}(\eta_1 kR_1)}={h^{(2)}_{l-1}(kR_1)}/{\eta_1 h^{(2)}_{l}(kR_1)};\\
    &\text{Mode $TM_{lms}$:\hspace{0.5cm}}\frac{l}{\eta_1 kR_1}-\frac{j_{l-1}(\eta_1 kR_1)}{j_{l}(\eta_1 kR_1)}=\frac{l}{kR_1}-\eta_1\frac{h^{(2)}_{l-1}(kR_1)}{h^{(2)}_{l}(kR_1)}\cdot
    \label{eq:char_eq}
\end{aligned}
\end{align}
Here $j_{l}(\cdot)$ and $h^{(2)}_{l}(\cdot)$ are the spherical Bessel function of the first kind and the spherical Hankel function of the second kind of the order $l$, and $k$ is a free space wavenumber. These equations are independent of the azimuthal mode number $m$ because of spherical symmetry. The latter implies that all $2l+1$ modes with different $m$ are all degenerate in frequency.

In the approximation $\eta_1 \gg 1$, the resulting zeros of the spherical Hankel function give solutions for the waves outside  the sphere ($kR_1$-dependent ``external'' modes), whereas the zeros of the spherical Bessel function give solutions for the waves inside the sphere ($\eta_1 kR_1$-dependent ``internal'' modes). For the ``external'' modes the energy is concentrated on the surface or outside the sphere. The energy of ``internal'' modes is concentrated in the entire volume of the sphere and on its surface, but with the predominance of the stored energy inside the sphere. 
Since the $Q$ factor of ``external'' modes is less than unity~\cite{Gastine1967} and they cannot be efficiently excited in electrically small antennas, we focused only on the ``internal'' modes which may have significantly higher $Q$ factor. The quality factor, $Q$, is a dimensionless parameter that can be roughly defined for leaky modes as the ratio of the energy stored in the resonator to the energy lost in one radian of the oscillation cycle.
\begin{wraptable}{r}{8.4cm}
\centering
\begin{tabular}{ c | c c c c }
 \hline
   $ $& 
   $s=1$  & 
   $s=2$  & 
   $s=3$  & 
   $s=4$  \\ [0.5ex]
 \hline
 $l=0$ & $3.14159$     & $6.28319$    & $9.42478$    & $12.5664$   \\
 $l=1$ & $4.49341$ & $7.72525$ & $10.9041$ & $14.0662$\\
 $l=2$ & $5.76346$ & $9.09501$ & $12.3229$ & $15.5146$\\
 $l=3$ & $6.98793$ & $10.4171$ & $13.698$  & $16.9236$\\
 $l=4$ & $8.18256$ & $11.7049$ & $15.0397$ & $18.3013$\\
 $l=5$ & $9.35581$ & $12.9665$ & $16.3547$ & $19.6532$\\
 $l=6$ & $10.5128$ & $14.2074$ & $17.648$  & $20.9835$\\
 \hline
\end{tabular}
\caption{\label{table:rescon}
Zeros $\zeta_{ls}$ of the spherical Bessel function of the first kind $j_{l}(\zeta_{ls}=\eta_1 k R_1)=0$. The subscript $s$ denotes here the ordinal number of the zero of the $l$-th order spherical Bessel function.}
\vspace{-1mm}
\end{wraptable}

Finally, the approximate resonance conditions for $TE_{(l+1)ms}$ and $TM_{lms}$ modes in the $\eta_1 \gg 1$ approximation:
\begin{equation}
    j_{l}(\eta_1 kR_1) \simeq 0
    \label{eq:res_cond}
\end{equation}
The first several zeros $\zeta_{ls}$ of $j_l(\zeta_{ls})=0$ are presented in Table \ref{table:rescon}.

\section{Results}
\label{sec:results}

\begin{figure}[b!]
\centering
\includegraphics[width=\textwidth]{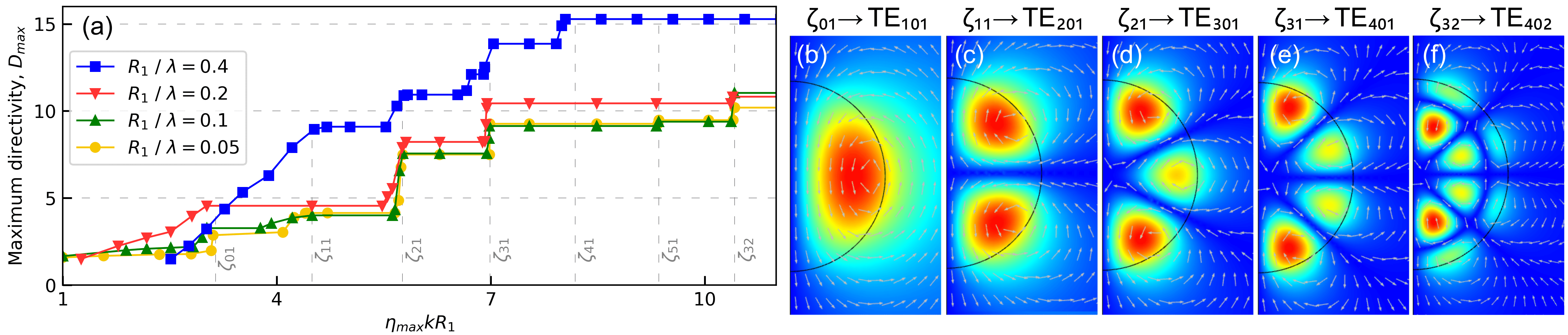}
\par
\caption{\label{fig:4Rl}
(a) Optimized directivity as a function of the $\eta_{max} k R_1$ value for antennas with a different $R_1/\lambda$ ratio for $1\leq\eta_{1}\leq\eta_{max}$. (b)-(f) A graphic representation of the modes of the sphere corresponding to different resonance conditions. The sphere modes are mirror-symmetrical, so only half of the sphere's cross section is demonstrated. The normal component of the electric field is shown in color; magnetic field lines are shown with gray arrows.}
\end{figure}
The directivity of an antenna is a measure of how much it concentrates power in a given direction. The maximum directivity of a homogeneous sphere in forward direction with an optimized refractive index $\eta_1 \leq \eta_{max}$ and dipole position $r_d$ is presented in Figure \ref{fig:4Rl}(a). With a decrease of the $R_1/\lambda$ ratio, the resonance behavior becomes stronger due to the weakening of the interference between the excited modes of the sphere. From equation (\ref{eq:res_cond}) it is known that the approximate conditions for the $TE_{(l+1)ms}$ and $TM_{lms}$ modes coincide. 
Thus the influence of these two modes affects the far-field scattering pattern at resonance condition and causes the jumps in directivity. If an electric dipole effectively excites resonances $\zeta_{ls}$ with $l \leq 3$ for a very small $R / \lambda$, then for subsequent $l$ it becomes ineffective and a further small gradual increase in directivity occurs with an increase in the index $s$. 
Figures \ref{fig:4Rl}(b)-(f) show the distribution of the electric and magnetic fields for the first $TE$-modes of the sphere. 
\begin{figure}
\begin{minipage}[c]{0.48\linewidth}
\includegraphics[width=\linewidth]{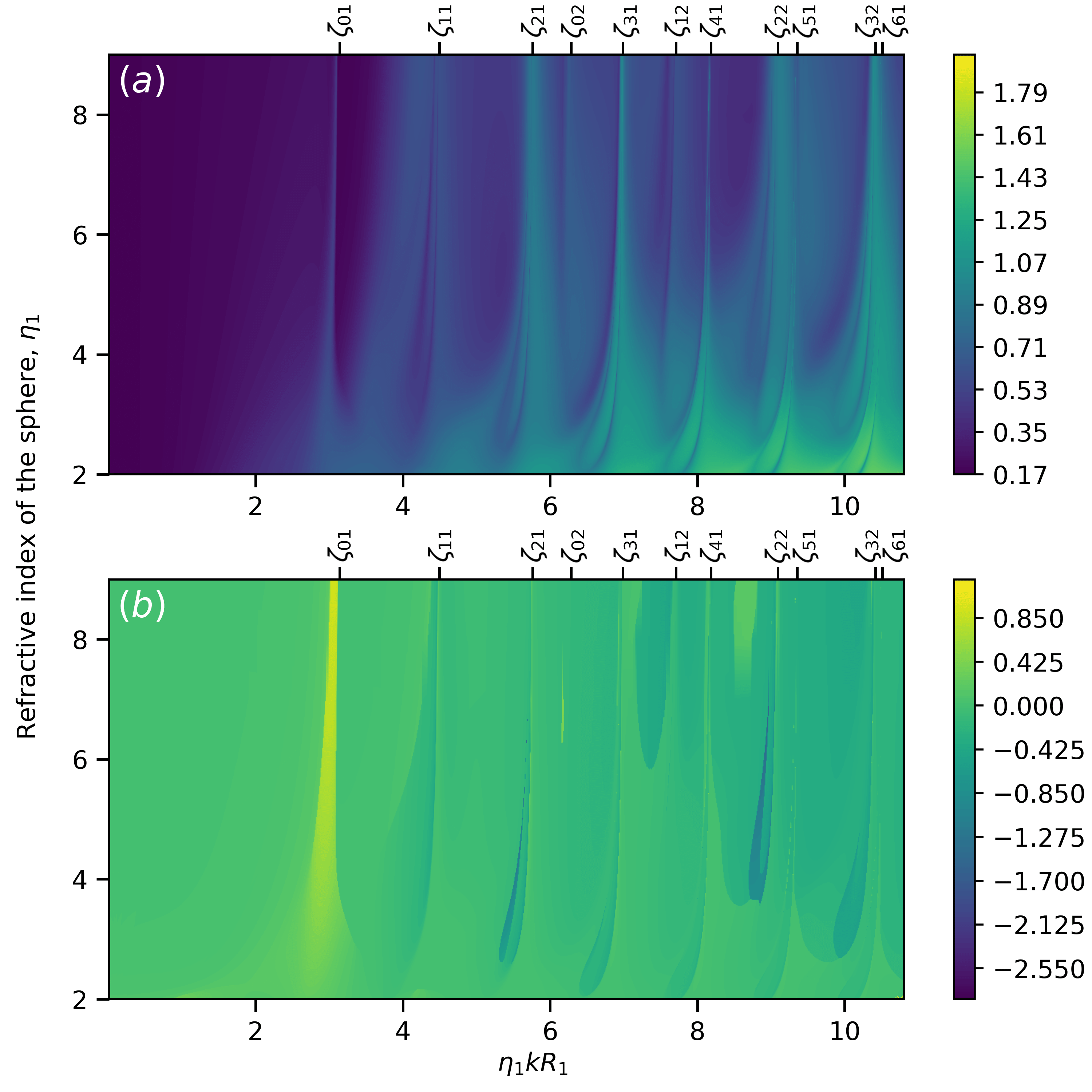}
\caption{\label{fig:dir_dip}
Optimized results for a homogeneous sphere on a logarithmic ($log_{10}$) scale: \mbox{(a) maximum directivity in forward direction}, \mbox{(b) relative dipole position, $r_d / R_1$}.}
\end{minipage}
\hfill
\begin{minipage}[c]{0.48\linewidth}
\includegraphics[width=\linewidth]{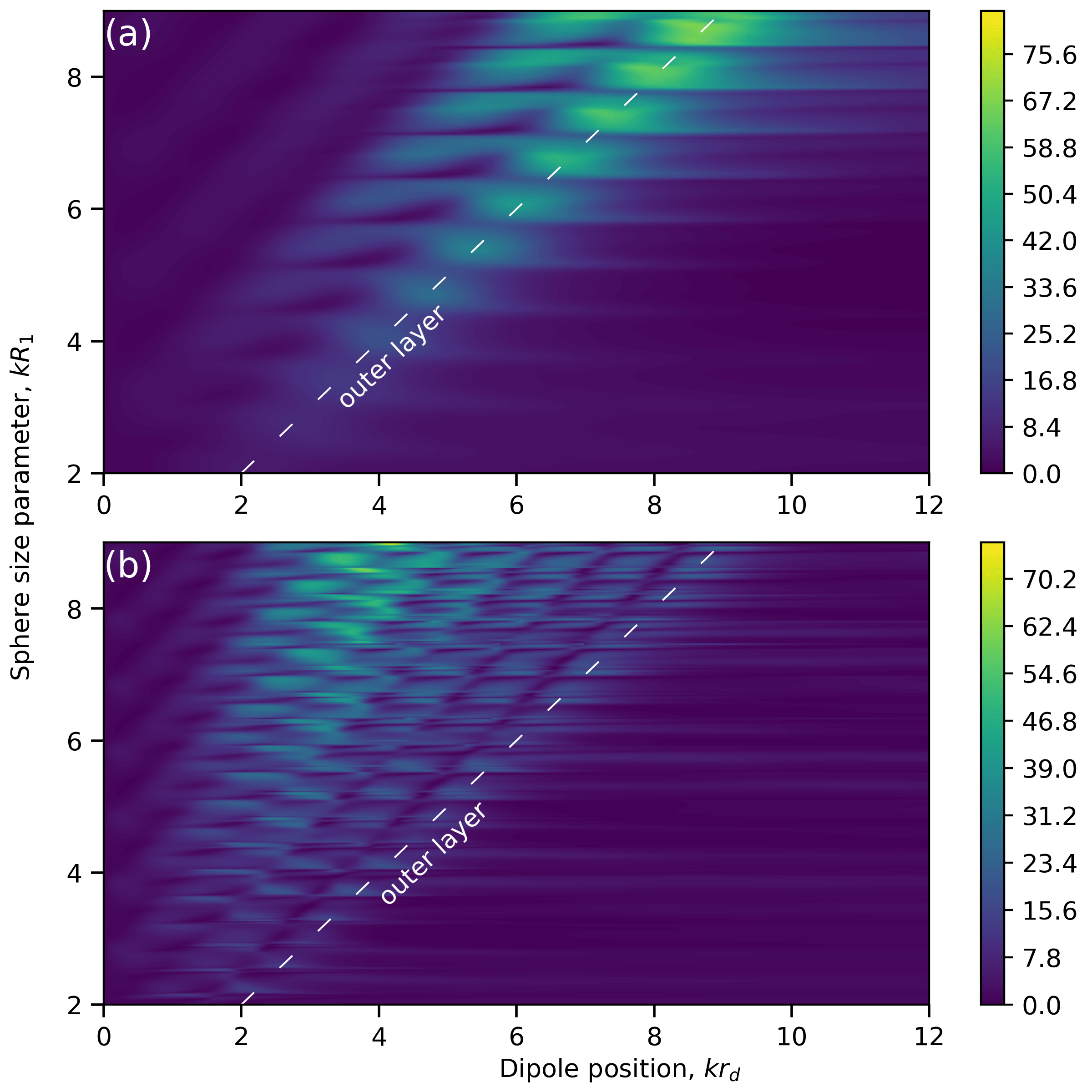}
\caption{\label{fig:3eps}
Dependence of the directivity of a homogeneous DRA on its geometric size and the position of the dipole: \mbox{(a) $\eta_1=\sqrt{3}$}, \mbox{(b) $\eta_1=\sqrt{10}$}.}
\end{minipage}%
\vspace{-1mm}
\end{figure}

The impact of interference between modes on the maximum directivity in forward direction at low refractive indices can be seen in Figure \ref{fig:dir_dip}(a), which shows the dependence of directivity on the refractive index $\eta_1$ and on the parameter $\eta_1 k R_1$. Figure \ref{fig:dir_dip}(a) also gives a visual representation of the accuracy of the conditions (\ref{eq:res_cond}) under the applied approximation for $\eta_1 \gg 1$. 
Moreover, while maintaining the value of $\eta_1 k R_1$ and decreasing $R_1$, the maximum directivity remains practically unchanged at all resonances.
Figure \ref{fig:dir_dip}(b) demonstrates the optimal dipole position which provides maximum directivity. If for the $\zeta_{01}$ resonance only mode $TE_{101}$ is excited and the dipole position $r_d$ tends to half the wavelength with increasing $\eta_1$, then for the remaining resonances the dipole position is either inside the sphere or near its surface, in order to excite two overlapping $TE_{(l+1)ms}$ and $TM_{lms}$ modes with proper amplitudes and phases. 

Figure \ref{fig:3eps} shows the dependence of the directivity on the physical size of a homogeneous sphere and on the position of the dipole for different refractive indices.
With an increase in the geometric size of the sphere, modes with a new index $s$ can appear in it, and this gives characteristic stripes in the figure. It becomes possible to excite such resonances from a larger number of dipole positions located inside the sphere ( $\sim s$ positions).

\section{Conclusions}
\label{sec:conclusions}

In this work we analyzed a directivity of an electrically small spherical DRA excited by a point electric dipole and its dependence on the effectively excited modes of the spherical DRA. 
We demonstrated the applicability of $j_{l}(\zeta_{ls} = \eta_1 kR_1) \simeq 0$ condition for $\eta_1 \gg 1$ regime to identify and characterize excited modes of a sphere.
Efficient excitation of the resonances $\zeta_{ls}$ is possible by a dipole located inside the sphere or in the immediate vicinity of its surface (for $l>0$). 
Maximum directivity of considered DRAs can be achieved at $\zeta_{3s}$ resonances.



\ack
This work is supported by the Russian Foundation for Basic Research (project \#20-32-90243). 

\bibliographystyle{unsrt}

\end{document}